\PassOptionsToClass{
reprint,
superscriptaddress,
amsmath, amssymb,
aps,
prb,
floatfix,
}{revtex4-2}
\documentclass{revtex4-2}

\usepackage{graphicx}%
\usepackage{dcolumn}%
\usepackage{bm}%
\usepackage{enumitem} %
\usepackage[colorlinks=true, allcolors=blue]{hyperref}%
\usepackage[normalem]{ulem} %

\usepackage{bibunits}
\defaultbibliography{bibliography}

\defaultbibliographystyle{apsrev4-2-title}

\usepackage[T1]{fontenc}
\usepackage{newtxtext,newtxmath}

\graphicspath{{figures/}}
\usepackage[separate-uncertainty=false]{siunitx}
\usepackage{braket}

\usepackage[capitalise]{cleveref}

\usepackage{changes}
\usepackage{xcolor}

\definechangesauthor[name={Dimitris}, color=orange]{dt}
\definechangesauthor[name={Antonio}, color=blue]{ag}
\definechangesauthor[name={Yannis}, color=green]{yg}
\definechangesauthor[name={Helgi}, color=purple]{hs}
\definechangesauthor[name={revisions}, color=black]{rev}

\newcommand{\rev}[1]{#1}

\begin{document}
\begin{bibunit}

\title{Geometric control of hyperbolic exciton-polariton condensate dimers}

\date{\today}

\author{Ioannis Georgakilas}
\altaffiliation{These two authors contributed equally.}
\affiliation{IBM Research–Zurich, S\"aumerstrasse 4, CH-8803 R\"uschlikon, Switzerland}
\affiliation{Institute of Quantum Electronics, ETH Zurich, Auguste-Piccard-Hof 1, 8093 Zürich, Switzerland}

\author{Antonio Gianfrate}
\altaffiliation{These two authors contributed equally.}
\affiliation{CNR Nanotec, Institute of Nanotechnology, via Monteroni, 73100, Lecce, Italy}

\author{Dimitrios Trypogeorgos}
\email[Corresponding Author: ]{dimitrios.trypogeorgos@cnr.it}
\affiliation{CNR Nanotec, Institute of Nanotechnology, via Monteroni, 73100, Lecce, Italy}

\author{Helgi Sigurðsson}
\affiliation{Institute of Experimental Physics, Faculty of Physics, University of Warsaw, ul. Pasteura 5, PL-02-093 Warsaw, Poland}
\affiliation{Science Institute, University of Iceland, Dunhagi 3, IS-107 Reykjavik, Iceland}

\author{Fabrizio Riminucci}
\affiliation{Molecular Foundry, Lawrence Berkeley National Laboratory, One Cyclotron Road, Berkeley, California, 94720, USA}

\author{Kirk W. Baldwin}
\author{Loren N. Pfeiffer}
\affiliation{PRISM, Princeton Institute for the Science and Technology of Materials, Princeton University, Princeton, New Jersey 08540, USA}

\author{Milena De Giorgi}
\author{Dario Ballarini}
\author{Daniele Sanvitto}
\affiliation{CNR Nanotec, Institute of Nanotechnology, via Monteroni, 73100, Lecce, Italy}

\begin{abstract}

Coupled many-body quantum systems exhibit rich emergent physics with diverse stationary and dynamical behaviors. By engineering platforms with tunable and distinct coupling mechanisms, new insights emerge into the collective behavior of coupled many body systems. Particles can be exchanged via evanescent or ballistic coupling: the former, based on proximity, yields large spectral splitting, while the latter requires strict phase-matching, analogous to phase-coupled harmonic oscillators, and has a smaller impact on the energy landscape. We demonstrate an all-optically tunable quantum fluid dimer based on exciton-polariton condensates in a photonic crystal waveguide with hyperbolic (saddle-like) dispersion. Varying the dimer's angle relative to the grating tunes the coupling from evanescent to ballistic. We directly observe spectral features and mass flow shaped by the saddle dispersion. This work highlights photonic crystals as powerful platforms to explore condensed matter phenomena lying at the interface between delay-coupled nonlinear oscillators and tight binding physics.

\end{abstract}

\maketitle
\noindent
\textbf{$^*$} These authors contributed equally \\
\textbf{$^\dagger$} Corresponding Author: Dimitrios Trypogeorgos, dimitrios.trypogeorgos@cnr.it

\section{Introduction}

From sub-atomic particles, to atoms, to molecules, our physical universe emerges from interacting elementary particles.
For systems that approach a mesoscopic size the details of their shared wavefunction depend on the nature of their coupling.
For the most part, a coupled phase-amplitude oscillator model is illustrative of the two possible mechanisms: quantum systems will either synchronise in phase or open spectral gaps and initiate dynamical oscillation when in proximity with each other.
These two fundamentally different mechanisms are responsible, on the one hand, for various synchronisation~\cite{RevModPhys.77.137} and ballistic transport~\cite{calado_ballistic2015} phenomena and, on the other hand, for the emergence of band structures in tight binding models~\cite{ashcroftSolidStatePhysics1976}, and super- and sub-radiance in arrays of quantum emitters~\cite{masson_natcomm2022,yu_sciadv.1701696}.
\rev{Continuous tuning between the nature of the coupling between these two regimes likely requires a hyperbolic dispersion, which naturally arises in systems with mixed periodic and continuous translation symmetry, a situation which can be engineered in artificial systems.
Here, we demonstrate how this is possible in a photonic crystal exciton-polariton waveguide, where the continuous symmetry stems from the planar geometry of the waveguide and the discrete one from the inscribed unidirectional grating.
}

Exciton-polaritons (polaritons from here on) are bosonic light-matter quasiparticles that form under the strong coupling of confined photons and quantum well excitons~\cite{Deng_RMP2010}.\rev{ They possess a small effective mass (stemming from the photonic component) and exhibit dipole-dipole interactions (driven by the excitonic component). These two properties, when combined with high-quality photonic modes, enable their nonequilibrium condensation at elevated temperatures across a wide range of materials~\cite{Ghosh_PhotIns2022}. } The coupling between neighbouring polariton condensates is defined by their engineered potential landscape, excitation technique, on-site energy, and separation distance which gives access to a variety of distinct polariton Hamiltonians~\cite{Amo_CRPhys2016}. As such, considerable effort has been dedicated towards controlling the environment and potential landscape and, consequently, the coupling between polariton condensates through intricate sample fabrication and patterning~\cite{Schneider_RPP2016} and all-optical techniques~\cite{Alyatkin_PRL2020, Tao_NatMat2022, Ohadi_PRL2017, Dovzhenko_PRB2023, gianfrateReconfigurableQuantumFluid2024,alyatkin_antiferromagnetic_2024,harrison_synchronization_2020}. 

To date, mode hybridization and population oscillations of deeply confined evanescently-coupled polariton condensates have been confirmed between accidental sample defects~\cite{Lagoudakis_PRL2010} or in pre-designed photonic micropillar molecules~\cite{Galbiati_PRL2012, Abbarchi_NatPhys2013}. 
There, the usual tight-binding approach, supplemented with non-Hermitian terms, captures the essential physics between coupled polariton condensates. In contrast to deeply confined condensates, there are ballistic polariton condensates which are defined by strong radial polariton outflow and stringent phase matching conditions between neighbours that defines their synchronisation~\cite{Ohadi_PRX2016, Topfer_Communications_2020, Alyatkin_PRL2020, Furman_CommPhys2023, Dovzhenko_PRB2023}. 
These ballistic condensates appear when pumped with a tightly focused beam, which leads to spatially localised blueshift and gain and strong out-of-equilibrium polariton behaviour. 

\rev{It is important to notice that these two coupling mechanisms cannot be straightforwardly described as purely Hermitian or anti-Hermitian. Consequently, they differ fundamentally from the well-studied dissipative and dispersive couplings \cite{ding_dispersive_2019,toebes_dispersive_2022}. This is because both mechanisms affect the real but also the imaginary components of the polariton condensates.
The more pronounced difference is observed in the ballistic coupling. Although it may initially appear to be predominantly dissipative, it actually involves a direct exchange of population leading to eigenvalues with a small but significant real gap and no indication of an imaginary gap.}

\begin{figure}[tbp]
\centering
\includegraphics[width=\columnwidth]{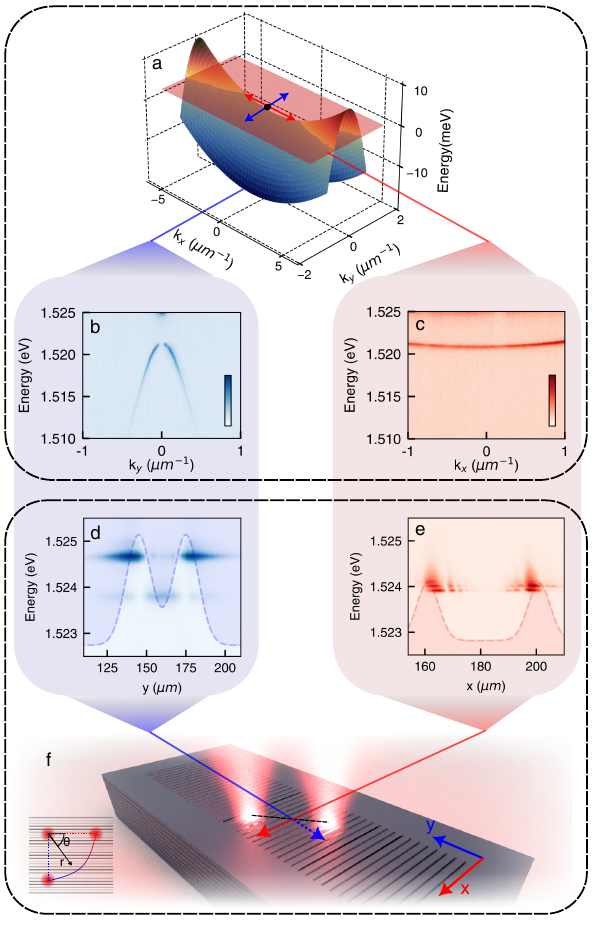}
\caption{\textbf{System and geometry.} 
\textbf{a}. 3D representation of the dispersion of the polaritonic waveguide. The lowest-energy mode, where condensation takes place, has a hyperbolic, saddle-like, dispersion which inverts the effective mass of the two principal reciprocal directions in the plane of the waveguide\rev{, the x direction refers to the direction parallel to the grating grooves and the y to the direction along the grating period}. \textbf{b} and \textbf{c}. Dispersions along $k_x$ and $k_y$ showing the different effective mass in the lower branch. \textbf{d} and \textbf{e}. The energy-resolved emission from a polariton dimer in coordinate space. Overlayed is the effective potential created by the pump spots that is attractive in \textbf{d} and repulsive in \textbf{e}. The potentials are drawn taking the effective mass into account, in such a way that the condensate forms in the non-coloured area. \textbf{f}. Schematic of the pump profiles on the grating waveguide. The coupling is ballistic (evanescent) along the $x$-axis ($y$-axis).
}
\label{fig:1}
\end{figure}

Here we propose a system of reduced symmetry, in contrast to cylindrically symmetric planar microcavities, to access a new geometric tuning parameter for the inter-condensate coupling mechanism. 
Our system is a subwavelength grated waveguiding slab with multiple embedded quantum wells [see schematic \cref{fig:1}] that supports optically reconfigurable bound-in-the-continuum (BiC) polariton condensates~\cite{ardizzonePolaritonBoseEinstein2022, gianfrateReconfigurableQuantumFluid2024}. 
The properties of BiC polaritons are notably different from conventional cavity polariton platforms. 
They exhibit extremely low radiative decay rates, \rev{which substantially decreases their condensation threshold}~\cite{ardizzonePolaritonBoseEinstein2022,riminucciPolaritonCondensationGapConfined2023, Wu2024} as compared to ballistically coupled condensates~\cite{Topfer_Communications_2020}.
\rev{Additionally, the low radiative decay enables the attainment of high densities suitable} for the study of highly nonlinear quantum hydrodynamics~\cite{Smirnov_PhysRevB.89.235310,Grosso_PhysRevLett.107.245301,trypogeorgos2024emergingsupersoliditypolaritoncondensate} that requires long spatio-temporal coherence scales~\cite{Fontaine_Nature_2022}. 
When dimerised, in this highly anisotropic photonic crystal, the angle of the axis connecting two condensates, with respect to the grating direction, plays a pivotal role in determining whether the coupling mechanism can be interpreted as evanescent or ballistic in origin, or a mixture of the two. 
The ability to continuously tune between the two natures of inter-condensate coupling introduces a new paradigm in polaritonic lattice systems, which until now have remained limited to either fully ballistic or fully evanescent coupling mechanisms. 
\rev{In the following, we connect the coexistence of these two coupling mechanisms with the geometrical properties of a repulsive polariton condensate lying on a hyperbolic dispersion. We directly visualize the effects of the self-confining mechanism and investigate its phenomenology and pronounced directionality.}

\begin{figure*}[htbp]
\centering
\includegraphics[width=\linewidth]{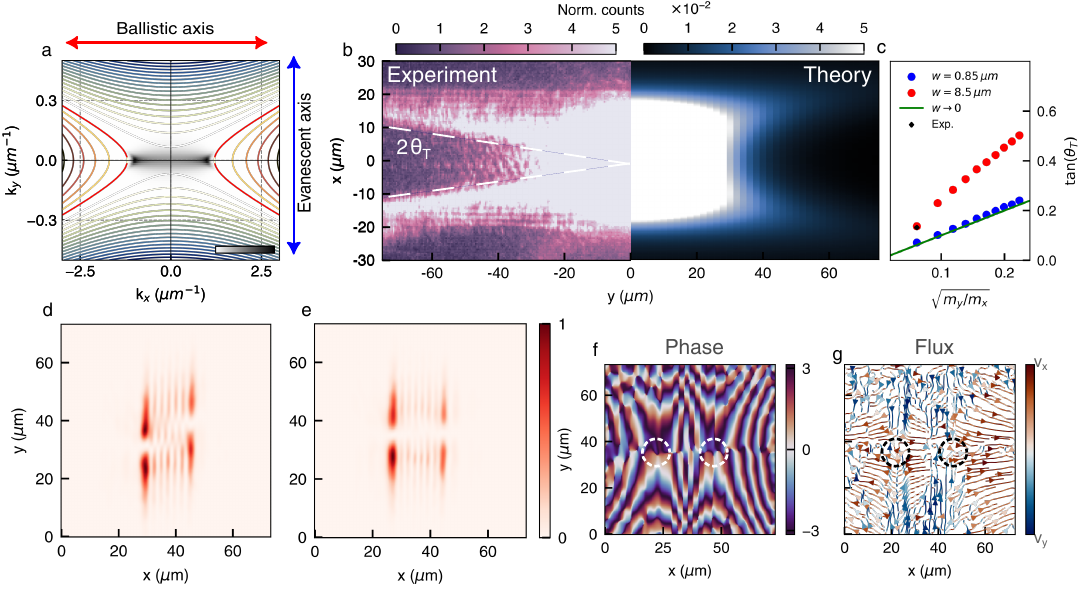}
\caption{\textbf{Polariton flow on hyperbolic contours.}
\textbf{a}. Iso-energetic curves at a blueshift of \qty{0.4}{meV} from the saddle point showing two opposite hyperbolas. Along $k_x$ the dispersion is convex parabolic leading to strong polariton outflow and ballistic coupling (red arrow), while along $k_y$ it is concave parabolic (negative effective mass) which leads to optical trapping and evanescent coupling (blue arrow) of the condensates. The calculated reciprocal-space PL for a single condensate on the saddle dispersion is superimposed. 
\textbf{b}. Oversaturated coordinate-space PL above condensation threshold. Since, along the y direction, there exist no available states for the polaritons to scatter to, propagation is not allowed within a cone of about \rev{\qty{\pm 7.5}{\degree}}, left experiment, right theory. \textbf{c.} Dependence of $\theta_T$ on the mass ratio and the \rev{ trap size $w$} from simulations. \textbf{d-e.} Coordinate-space PL of two condensates at different angles in the ballistic regime. The fringes show high directionality and are always oriented along $y$. \textbf{f-g}. The overall flow is better visualised from the phase and velocity images showing inflow (outflow) of polaritons along the $x$-axis ($y$-axis). The red-blue colorscale of the velocity arrows denotes their directionality as $\left| \arctan(v_x/v_y) \right|$.}
\label{fig:2}
\end{figure*}

\section{Results}

\subsection{Evanescent and ballistic axis}
The studied waveguide has a two-dimensional (2D) slab geometry, with the grating along one direction ($y$) and quasicontinuous translation symmetry along the other ($x$). Here, the lowest energy polariton branch forms a saddle dispersion centred at $k=0$ [see \cref{fig:1}a] where $\mathbf{k} = (k_x,k_y)^\text{T}$,
\begin{equation}
    \hat{\epsilon}(\mathbf{k}) - \epsilon_0 \approx \frac{\hbar^2}{2}\left( \frac{k_x^2}{m_x}-\frac{k^2_y}{m_y}\right).
    \label{eq:dispersion}
\end{equation}
This dispersion relation is in sharp contrast to typical planar polariton cavities which are described by a paraboloid, $\hat{\epsilon}(\mathbf{k})_\text{par} = \hbar^2 k^2/2m$ due to cylindrical symmetry. Here, $\epsilon_0$ is the energy of the lower branch polaritons at normal incidence and $m_{x,y}$ denote the magnitude of the effective masses along each direction. Along the grating direction, $\hat{\mathbf{y}}$, lower energy antisymmetric polaritons obtain a large negative effective mass [see \cref{fig:1}b] because guided photon modes fold across the Brillouin zone and open a gap at $k_y=0$~\cite{Sigurdsson_NanoPho2024}, whereas in the perpendicular direction, $\hat{\mathbf{x}}$, \rev{ the photons propagating parallel to the grating-groves direction retain their positive effective mass.} [see \cref{fig:1}c] from standard photon confinement along the $z$-direction. More details on the full polariton dispersion is shown in Supplementary Fig.~S1.

The anisotropy of the saddle dispersion gives a marked directionality in the coupling mechanism between two neighbouring condensates, driven by two nonresonant pump spots (Gaussian beams) focused on the grating [see \cref{fig:1}f],
\begin{equation}
    P(\mathbf{r}) = e^{-|\mathbf{r}-\mathbf{r}_1|^2/2w^2} + e^{-|\mathbf{r}-\mathbf{r}_2|^2/2w^2}.
\end{equation}
The separation distance between the two condensates is denoted by $r = |\mathbf{r}_1-\mathbf{r}_2|$ and the angle of their link with respect to the $x$-axis as $\theta$. The nonresonant pumps not only sustain the condensates above threshold, but also photoexcite an incoherent background of charge carriers and excitons which, due to the strong Coulomb interactions of excitons, results in large polariton blueshifts. In this sense, the pump profile is proportional to an effective potential landscape $P(\mathbf{r}) \propto V(\mathbf{r})>0$ felt by the polaritons~\cite{gianfrateReconfigurableQuantumFluid2024}. Because of the pump-induced potential gradient, positive and negative effective mass polaritons, are 'repelled from' and 'attracted to' their pumped regions respectively [see \cref{fig:1}d,e] where the potentials have been drawn taking into account the difference in the effective mass sign, ensuring that the condensate forms in the unshaded region.

Each scenario displays qualitatively different behaviour. Negative-mass polaritons experience the two pumps as a double potential well that traps them deeply [see \cref{fig:1}d]. On the other hand, positive mass polaritons experience the two pumps as repulsive potentials, that may instead trap particles between them when their separation is small enough [see \cref{fig:1}e]~\cite{Tosi_NatPhys2012, Ohadi_PRX2016}. For this reason, the two directional extremes $\theta = 0^\circ$ and $90^\circ$ are referred to as the {\it ballistic} and {\it evanescent} coupling directions, respectively, corresponding to the dominant coupling mechanism. Along the evanescent $y$-direction, slowly varying condensed polaritons tunnel through a forbidden region resulting in mode hybridization with a consequent bonding and an antibonding level \cref{fig:1}d separated by $\sim$meVs in energy~\cite{gianfrateReconfigurableQuantumFluid2024}.
Orienting the two pump spots along the ballistic $x$-direction, however, condensed polaritons couple to propagating modes between the pump spots resulting in a so-called ballistic coupling mechanism. Here, the two pump spots form a double-barrier potential along the ballistic $x$-axis with propagating positive-mass polaritons forming many standing wave resonances whose smaller energy spacing [see \cref{fig:1}e] is given by the distance between pump spots, similar to results reported for in planar cavities~\cite{Topfer_Communications_2020}.

\begin{figure*}[htbp]
\centering
\includegraphics[width=2\columnwidth]{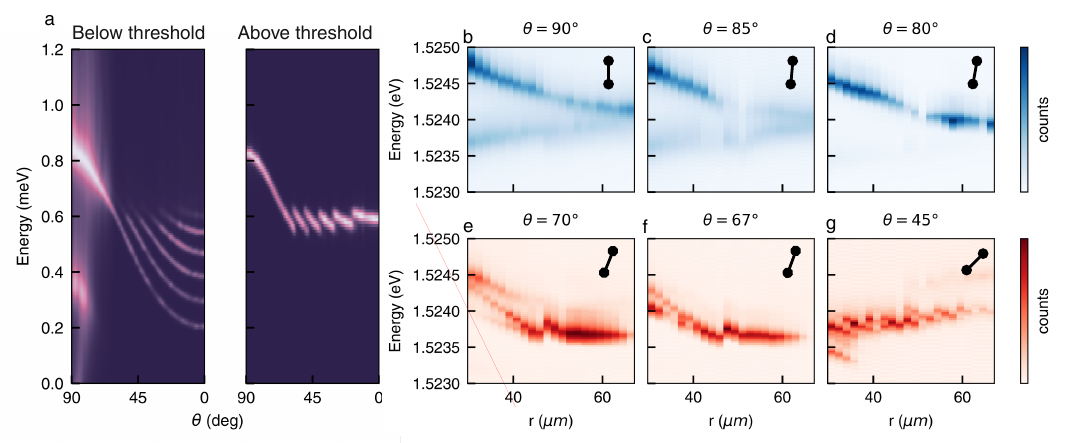}
\caption{\textbf{Transition between evanescent and ballistic coupling regimes.} \textbf{a} Calculated spectrum of the polariton dimer over angle for a fixed distance of \qty{30}{\mu m} (left) below condensation threshold and (right) $1\%$ above threshold. \textbf{b-g} PL showing the energy spectrum over distance $r$, for various angles $\theta$. The energy gap of the dimer decreases by increasing distance. When entering the ballistic regime \textbf{e-g} we observe the characteristic mode-flipping between the in-phase and anti-phase configurations.
The energy for $r \gg 1$ asymptotically goes to that of a single condensate; slight variations of the order of \qty{300}{\micro eV} are due to 5\% fluctuations in the laser power.
}
\label{fig:3}
\end{figure*}

The observed directional evanescent and ballistic coupling mechanisms can be appreciated from the opposite-facing isofrequency hyperbolas in the polariton saddle dispersion relation [see \cref{fig:2}a],  
\begin{equation}
    k_y =  \pm \sqrt{ m_y \left(\frac{k_x^2}{m_x} - \frac{2\epsilon_c}{\hbar^2} \right)}.
\end{equation}
The aspect ratio of these hyperbolas is determined by the effective mass ratio $m_x/m_y$ that, in our system, is $\approx 263$ by comparing \cref{fig:1}b and \cref{fig:1}c at small momenta. 
Here, we have set $\epsilon_0=0$ without loss of generality, and $\epsilon_c>0$ is the energy of the condensate which is set by the laser power (pump-induced blueshift). 
Condensed polaritons populate a pair of opposite-facing hyperbolic contours in the dispersion, which results in strong anisotropic PL (density) distribution for the single condensate shown in \cref{fig:2}b with an oversaturated colorscale. 
Along the evanescent $y$-direction the polaritons find no available states to scatter to, whereas along the ballistic $x$-direction their outward flow is possible but with a much heavier mass. 
This yields a polariton ``propagation cone'' separating a dark forbidden region from a brighter ballistic region. 
The angle of the real-space cone $\theta_T$ is reciprocally related to the slopes of the populated hyperbolic contours, which, in the asymptotic limit $k \to \infty$ is $s = \pm \sqrt{m_y/m_x}$. 
Therefore, for a small pump spot that excites large-wavenumber polaritons, we obtain $\theta_T \to \arctan{(\sqrt{m_y/m_x})} = \arctan{(\sqrt{1/263})} = \ang{3.5}$. In experiment, the finite width of the pump spot creates an optical trap of approximately w $\approx 8.5$ $\mu$m which causes redistribution of the polariton condensate in momentum space, affecting the angle $\theta_T$. The experimentally measured angle of the forbidden region from Fig.~\ref{fig:2}b is $\theta_T \approx \ang{7.5}$ which is in good agreement with the angle extracted from mean field simulations [see SI for details] shown in Fig.~\ref{fig:2}c.
We also show the numerically calculated $\theta_T$ for a narrow trap of $\text{w} = \qty{0.85}{\micro m}$. 
The corresponding calculated reciprocal and real space condensate density for a trap (exciton reservoir) of w $ = \qty{8.5}{\micro m}$ is shown in \cref{fig:2}a, overlapped with the dispersion contours, and in \cref{fig:2}b, respectively. Notice the two maxima at $k_c \approx \pm \qty{1}{\micro m^{-1}}$ which define the wavenumber of ballistic polariton waves along the $x$-direction.

Figures~\ref{fig:2}d,e show example coordinate space PL distributions of the condensate dimer for two different angles $\theta = 26^\circ$ and $0^\circ$. 
We note that the central nodal line in each condensate is a property of the BiC state~\cite{ardizzonePolaritonBoseEinstein2022}. 
Between the condensates multiple interference fringes appear due to positive-mass propagating waves reflecting back-and-forth between the pump spots. 
As the projection of the dimer link on the ballistic $x$-axis reduces with larger angle, $r_x = r \cos{(\theta)}$, the number of interference fringes between the spots also reduces. 
This can be seen in \cref{fig:2}d where the number of fringes are 5 whereas in \cref{fig:2}e they are 6. 
Namely, $r_x$ is reduced by $\lambda_c/2$, where $\lambda_c = \sqrt{2 \pi^2 \hbar^2 / m_x \epsilon_c}$ is the average wavelength of the ballistic component of the condensates. 
Analogously, the condensates can be said to have undergone a transition from an anti-phase synchronous (even number of fringes) to an in-phase synchronous (odd number of fringes) state. 
As the dimer angle $\theta$ is increased further, the condensate dimer passes through more synchrony transitions until it enters the evanescent regime~\cite{gianfrateReconfigurableQuantumFluid2024}. 

\cref{fig:2}f and~\ref{fig:2}g show the extracted condensate dimer phase and velocity maps corresponding to \cref{fig:2}e. 
Positive-mass polaritons obtain an outward velocity component along the $x$-direction with respect to their pump spots, whereas along the $y$-axis negative-mass polaritons acquire an inward velocity component and are attracted to their pump spots.

\subsection{Hybrid directional coupling}

The angular and radial dependence of the coupling mechanism between the two condensates can be appreciated from their overlap integral which as a function of separation distance $\mathbf{r}_1 - \mathbf{r}_2$.
Using an ansatz composed of propagating and evanescent waves along each direction [see SI] we obtain the following,
\begin{equation} \label{eq:J}
\begin{split}
    J(r,\theta) & \propto \int \psi(\mathbf{r}_1)^* \psi(\mathbf{r}_2) \, d\mathbf{r}, \\
    & =  \cos{(k_c  r_x)} e^{-\kappa r_x} e^{-(r^2 - r_x^2)  / 8 \sigma_y^2}.
    \end{split}
\end{equation}
Here, $k_c = 2\pi/\lambda_c$ is the average wavenumber of the condensate outgoing waves, $\kappa \sim \gamma m_x/2 \hbar k_c$ is a damping coefficient due to the finite polariton lifetime $\gamma^{-1}$ that defines its extension along the ballistic $x$-direction, and $\sigma_y$ defines the size of the trapped condensate along the evanescent $y$-direction. 
The number of interference fringes between the condensates is associated with the $n$th root of the cosine function~\cite{Ohadi_PRX2016}. 
For $J>0$ the condensates synchronise in-phase whereas for $J<0$ they synchronize anti-phase in order to maximise the gain and amplitude of the dimer. It is important to note that $J$ does not completely describe the energy structure of the ballistic coupling which is instead determined by solutions of a transcendental problem of scattered polariton waves between the pump spots under open boundary conditions~\cite{Topfer_Communications_2020} [see SI for details, Supplementary Fig.~S2 and S3].

Here we focus on the optically tunable transition between the two coupling regimes. 
\cref{fig:3}a shows the calculated polariton spectral density for two pump spots as a function of $\theta$, below and above condensation threshold. Our calculations are based on the scalar 2D generalized Gross-Pitaevskii equation~\cite{Sigurdsson_NanoPho2024}, which can be regarded as a limiting case of nonlinear Maxwell-Bloch equations (see SI for details).
For a dimer orientated along the evanescent axis, $\theta=\ang{90}$, the spectrum reveals the bonding (upper) and antibonding (lower) branches~\cite{gianfrateReconfigurableQuantumFluid2024}. 
The reason the bonding branch is higher in energy, as opposed to being lower like in atomic orbital theory, is because energy-level hierarchy inverts when the effective mass is negative. 
As $\theta$ is decreased, the two major branches redshift revealing the appearance of additional branches, above $>\ang{45}$, corresponding to the resonances of the ballistic polariton dimer~\cite{Tosi_NatPhys2012, Ohadi_PRX2016, Topfer_Communications_2020}. 
Above threshold, the sharp spectral density of the condensate appears to jump from one resonance branch to the next when sweeping $\theta$, signifying transitions from in-phase to anti-phase synchronization, and so on~\cite{Ohadi_PRX2016, Topfer_Communications_2020}.

\begin{figure}[h!]
\centering
\includegraphics[width=\columnwidth]{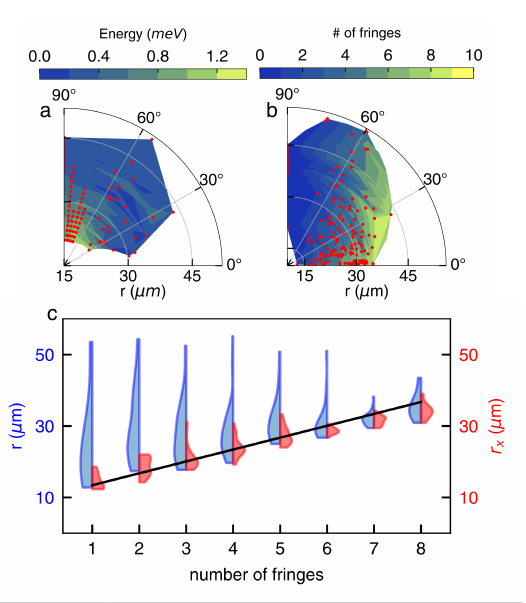}
\caption{\textbf{Directional coupling}. \rev{\textbf{a}. Energy splitting in the condensate dimer as a function of angle $\theta$ and distance $r$. \textbf{b}. Number of interference fringes parametrized as \textbf{b}.}
. The evanescent coupling decreases monotonically with both $\theta$ and $r$, ranging from \qtyrange{1.4}{0.1}{meV}. The ballistic coupling changes for in- to anti-phase with $r_x$, however it remains largely constant as long as $\theta < \ang{45}$. 
The red points indicate the coordinates of the measurements we used to extract the two couplings. The two colorbars indicate the width of the energy gap \textbf{a} and the number of fringes \textbf{b}. \textbf{c}. Statistical averages of the number of fringes for increasing distance between the condensates plotted as a function of $r$ and $r_x$. The width of the distribution is significantly smaller for $r_x$ showcasing the directionality of the system. The black line denotes the linear phase accumulation $k_c r_x/\pi$ of waves propagating between the condensates.
}
\label{fig:4}
\end{figure}

Corresponding experimental images are shown in \cref{fig:3}b-g as a function of separation distance $r$ and angle $\theta$, between \qtyrange{90}{45}{\degree} since for $\theta < \ang{45}$ the system is well into the ballistic regime and its spectral signature does not change significantly.
In the evanescent regime, the splitting between the bonded and the antibonded branches decreases as $e^{-r^2}$, as shown in \cref{fig:3}b-d, with polaritons becoming largely non-interacting at distances larger than $r>\qty{100}{\mu m}$. 
For small separations of $r<\qty{30}{\mu m}$, so that the two optical traps do not overlap significantly, we measured a considerable spectral gap of \qty{1.1}{meV} at $\theta = \ang{90}$ which is one order of magnitude larger than the polariton linewidth of \qty{100}{\mu eV}.
The trap overlap sets the maximum spectral gap attainable which decreases as $e^{-\sin^2{\theta}}$ going towards the ballistic regime.
This spatial dependence of $J$ can be seen clearly in \cref{fig:4}a.

As $\theta$ decreases, going from the evanescent to the ballistic regime, shown in \cref{fig:3}e-g, the bonding branch shows a gradual redshift while at the same time the antibonding branch gradually disappears, in agreement with theory shown \cref{fig:3}a.  
Close to \ang{70}, \cref{fig:3}e-f, the spectra starts showing characteristics of both coupling types; a considerable blueshift \rev{of the bonding state} for distances lower then \qty{40}{\mu m}, due to the condensate wavefunction overlap along the evanescent axis, together with the appearance of a substructure with an energy splitting in the order of \qty{100}{\mu eV} when in the ballistic coupling regime. 
Similar features can be identified in coordinate space, by looking at the characteristic magnitude of the energy splitting between distinct modes and the number of interference fringes between the condensates in their spatial and reciprocal space PL profile [see \cref{fig:1}e, Supplementary Fig.~S4 and S5.

\cref{fig:4}b shows the number of fringes ranging from \rev{1 to 9}. The increasing number of fringes with the spatial separation at small angles is similar to ballistic coupled condensates in microcavities~\cite{Topfer_Communications_2020}.
A new fringe enters the inter-condensate region every time the distance increases by more than an integer multiple of the fringe separation $\lambda_c/2$.
Differently from planar microcavities however, for a hyperbolic dispersion the outward ballistic polariton flow from each condensate is bounded by an in-plane ``propagation-cone'', forming a polariton jet of propagating waves with well-defined in-plane momenta [see \cref{fig:2}b]. If the two condensates are within each-other's cone aperture, ballistic coupling ensues. 
\cref{fig:4}c, shows the observed number of fringes as a function of either the radial distance $r$ or projected distance $r_x$. 
While the number of fringes is weakly dependent on $r$, the data distribution shows a clear linear trend as a function of $r_x$ with a slope $k_c r_x/\pi$ according to \cref{eq:J}, corresponding to phase accumulation of propagating ballistic waves. The extracted wavelength of outflowing condensate waves is $2\pi/k_c = \lambda_c\approx \qty{6.6}{\mu m}$ which is in good agreement with the two maxima in the calculated momentum space PL in \cref{fig:2}a (black intensity maxima).

\section{Discussion}

We have engineered a polaritonic platform based on a photonic crystal waveguide that exhibits the simultaneous presence of both evanescent and ballistic couplings and can smoothly transition between the two.
The nature of the coupling can be fully controlled geometrically by varying the angle connecting the polariton dimers.
These unique spatial features originate from the underlying hyperbolic dispersion of the system that we measured using reciprocal-space imaging. 
A parabolic dispersion along $k_x$ and a negative effective mass along $k_y$ lead to the ballistic and evanescent coupling of polaritons respectively. 
We provided a corresponding picture of the physics of BiC polariton condensates in coordinate space in terms of localisation geometry and in-plane fluxes. 
We showed that within a coordinate space circular sector, ballistic propagation is forbidden. 
Conversely, along the transverse direction, polariton propagation is characterised by a unidirectional flow with well-defined momenta. 
At the critical angle where the propagation cone appears the coupling transitions between the two variants cross a region where both types of couplings coexist.

We demonstrated reconfigurable formation of polariton dimers for a variety of distances and angles, fully mapping the positive quadrant.
The knowledge of the spatial dependence of the coupling is crucial for designing of complex networks of polaritons. 
Driven-dissipative networks of exciton-polariton Bose-Einstein condensates form a promising nonlinear optical platform for simulation of artificial lattices~\cite{Amo_CRPhys2016, Ohadi_PRL2017, Alyatkin_NatureComm_2021} and spin Hamiltonians~\cite{berloff.natmat.2017, Tao_NatMat2022}, studying universal scaling laws~\cite{Fontaine_Nature_2022}, topological physics~\cite{Solnyshkov_OME2021}, and implementing neural-inspired optical computation protocols~\cite{Opala_OME2023}. 
Our system goes beyond what is currently possible in the implementation of such networks given that the nature of the coupling parameter displays a highly anisotropic dependence on angle.
As such, our dimers constitute an \rev{novel} building component of advanced analogue polaritons simulators.

Insofar, we have ignored the non-trivial BiC topology, however it adds an additional powerful tool for engineering the coupling of system without depending exclusively on the changing effective mass.
The topology in coordinate space manifests as a $\pi$ phase difference between the two lobes which can lead to interesting phase structures depending on the position of the pumping sites.
The highly directional flow of polaritons that ensues between two pumping spots at a finite angle leads to their phase locking and at the same time develops a linear phase dislocation in the inter-condensate region; a single lobe of any given phase cannot satisfy the phase constraints when interacting with both lobes of another BiC that have a $\pi$ phase jump.
This attests to how topology can be used to enrich the physics of the coupled dimers even further.
Reducing the underlying symmetry of the system from $C_2$ to $C_4$, by imprinting a two-dimensional grating, effectively gives the same behaviour but at half-angle which can lead to tighter arrangement of pumping sites and more compact systems; however, note that the fundamental mechanism is perfectly contained in our one-dimensional waveguide.
Our platform provides a foundation for other studies to build upon and suggests novel paths of investigation using high-field trapped polaritons in a tight-binding setting.

\section{Methods}
Our system is a \qty{500}{nm} thick AlGaAs slab waveguide with 12 embedded \qty{20}{nm} GaAs quantum wells. 
The surface of the slab hetero-structure is etched with a \qty{170}{nm} deep grating and a \qty{243}{nm} period with the purpose of optimizing the exciton-photon detuning (slightly negative) and the diffractive coupling between counterpropagating waveguided modes [see \cref{fig:1}a]. 
The fabrication procedure is much simpler when compared to standard distributed Bragg reflector planar cavities, and is described in more detail in \cite{ardizzonePolaritonBoseEinstein2022,gianfrateReconfigurableQuantumFluid2024,riminucciPolaritonCondensationGapConfined2023}. 
The exciton-photon detuning is roughly $\Delta = -\qty{2.7}{meV}$, with a Rabi coupling strength of $\Omega = \qty{5.3}{meV}$. 
During the experiments the sample was held in a closed-loop helium cryostat at a temperature of \qty{18}{K}. 
Polariton condensates were excited using off-resonant continuous-wave excitation beam with wavelength of \qty{785}{nm} modulated at \qty{350}{Hz} with an optical chopper with 10\% duty cycle. 
A spatial light modulator was used to structure the beam into two Gaussian pump spots, each of size \qty{3}{\micro m} FWHM, the resulting luminescence i.e. the trap size $W$ is approximately \qty{20}{\mu m} FWHM.  %
The spot size was mainly limited by the numerical aperture of the excitation oblective.

\smallskip

\textbf{Data Availability.} 
The data of this study are available at https://doi.org/10.5281/zenodo.16911698 (ref. \cite{Data_repository}) Raw data can be obtained from the corresponding author on request.

\textbf{Code availability}
The code used in this study is available from the corresponding author upon request.

\smallskip

\onecolumngrid
\section*{References}
\twocolumngrid



%

\smallskip
\textbf{Acknowledgements.} 
We acknowledge fruitful discussions with R. Mahrt and T. St\"oferle.
This project was funded by PNRR MUR project: `National Quantum Science and Technology Institute' - NQSTI (PE0000023);
PNRR MUR project: ‘Integrated Infrastructure Initiative in Photonic and Quantum Sciences’ - I-PHOQS (IR0000016);
EU H2020 MSCA-ITN project “AppQInfo” (Grant Agreement No. 956071);
Quantum Optical Networks based on Exciton-polaritons - (Q-ONE) funding from the HORIZON-EIC-2022-PATHFINDER CHALLENGES EU programme under grant agreement No. 101115575;
Neuromorphic Polariton Accelerator - (PolArt) funding from the Horizon-EIC-2023-Pathfinder Open EU programme under grant agreement No.  101130304;
the project ``Hardware implementation of a polariton neural network for neuromorphic computing'' – Joint Bilateral Agreement CNR-RFBR (Russian Foundation for Basic Research) – Triennal Program 2021–2023;
the MAECI project ``Novel photonic platform for neuromorphic computing'', Joint Bilateral Project Italia - Polonia 2022-2023;  
the PRIN project ``QNoRM: A quantum neuromorphic recognition machine of quantum states'' - (grant 20229J8Z4P).
Views and opinions expressed are however those of the author(s) only and do not necessarily reflect those of the European Union or European Innovation Council and SMEs Executive Agency (EISMEA). Neither the European Union nor the granting authority can be held responsible for them.
This research is funded in part by the Gordon and Betty Moore Foundation’s EPiQS Initiative, grant GBMF9615 to L.P., and by the National Science Foundation MRSEC grant DMR 2011750 to Princeton University. 
Work at the Molecular Foundry is supported by the Office of Science, Office of Basic Energy Sciences, of the U.S. Department of Energy under Contract No. DE-AC02-05CH11231. 
We thank Scott Dhuey for assistance with electron beam lithography and Paolo Cazzato for the technical support.
H.S. acknowledges the Icelandic Research Fund (Rann\'{i}s), grant No. 239552-051, and the project No. 2022/45/P/ST3/00467 co-funded by the Polish National Science Centre and the European Union Framework Programme for Research and Innovation Horizon 2020 under the Marie Skłodowska-Curie grant agreement No. 945339.
\smallskip

\textbf{Author contributions.} 
IG and AG took the data and together with DT performed the analysis. 
AG, HS and DT wrote the manuscript. 
HS led the theory and performed all calculations.
FR processed the sample; growth was performed by KB and LP.
MDG, DB and DS supervised the work and discussed the data.
All authors contributed to discussions and editing of the manuscript.

\smallskip
\textbf{Competing interests.} 
The authors declare no competing interests.

\clearpage
\newpage

\end{bibunit}

\clearpage
\newpage

\begin{bibunit}

\end{bibunit}

\end{document}